\documentclass[submission,11pt]{eptcs}
\providecommand{\event}{HSB 2012} 

\usepackage[pdftex]{graphicx}
\usepackage{amsmath,amssymb,latexsym}

\usepackage{algorithmic}
\usepackage[ruled]{algorithm}
\usepackage{wrapfig}

\def\vy{\mathbf{y}}
\def\vp{\mathbf{p}}

\def\vc{\mathbf{c}}

\def\vx{\mathbf{x}}
\def\vg{\mathbf{g}}

\def\vz{\mathbf{z}}

\def\vva{\mathbf{a}}
\def\vvi{\mathbf{i}}
\def\vvd{\mathbf{d}}
\def\vvj{\mathbf{j}}

\def\vvv{\mathbf{v}}
\def\vvb{\mathbf{b}}

\def\Bern{\mathcal{B}}
\def\bern{\beta}

\def\polynom{\mathcal{\pi}}

\def\blossom{w}

\def\uBox{\mathcal{B}}

\def\transf{\tau}

\def\comp{\gamma}

\def\real{\mathbb{R}}

\DeclareMathAlphabet{\matheus}{U}{eus}{m}{n}
\DeclareMathAlphabet{\matheuf}{U}{euf}{m}{n}
\DeclareMathAlphabet{\mathsf}{OT1}{cmss}{m}{sl}
\DeclareMathAlphabet{\mathem}{OT1}{cmss}{m}{it}
\DeclareMathAlphabet{\mathgo}{U}{yswab}{m}{n}
\DeclareMathAlphabet{\mathcmr}{OT1}{cmr}{bx}{it}

\def    \s      {\sigma}

\def    \to     {\rightarrow}

\def    \se     {\subseteq}

\def    \lcc#1  {\langle\!\langle #1 \rangle\!\rangle}
\def    \rcc#1  {\langle #1 \rangle}
\def    \cc#1   {[#1]}

\def    \ft#1   {\footnote{#1} }

\def	\nodeq#1 {\stackrel{#1}{\sim}}
\def	\trans#1 {\stackrel{#1}{\longrightarrow}}
\def	\lead#1 {\stackrel{#1}{\leadsto}}

\def	\comment#1 {}

\def    \comm#1  {\begin{center} {\tt $<<<$ #1 $>>>$ } \end{center}}

\def \qed {\hfill\rule{2mm}{2mm}}

\def\GF{\hbox{\raise 1.6pt \hbox{$\varphi$}}}

\newcommand\conv{conv}

\renewcommand\trans[1]{\stackrel{#1}{\longrightarrow}}
\renewcommand\trans[2]{\longrightarrow\!\!\!\!\stackrel{#1}{#2}}
\renewcommand\trans[2]{\stackrel{#1}{\longrightarrow}_{_{\!\!\!\!\!\!\!\!\!#2 \ }}}

\def\real{\mathbb{R}}  

\newtheorem{claim}{Claim}
\newtheorem{lemma}[claim]{Lemma}
\newtheorem{theorem}{Theorem}

\title{Analysis of parametric biological models with non-linear dynamics}
\author{Romain Testylier
\institute{VERIMAG, Joseph Fourier University\\ Grenoble, France}
\thanks{This work is supported by the ANR project VEDECY.}
\email{Romain.Testylier@imag.fr}
\and
Thao Dang
\institute{CNRS/VERIMAG\\
France}
\email{Thao.Dang@imag.fr}
}
\def\titlerunning{Analysis of parametric biological models with non-linear dynamics}
\def\authorrunning{Romain Testylier \& Thao Dang}
\begin{document}
\maketitle

\begin{abstract}
In this paper we present recent results on parametric analysis of biological models. The underlying method is based on the algorithms for computing trajectory sets of hybrid systems with polynomial dynamics. The method is then applied to two case studies of biological systems: one is a cardiac cell model for studying the conditions for cardiac abnormalities, and the second is a model of insect nest-site choice. 
\end{abstract}

\section{Introduction}

Computer aided modelling and analysis have been intensively used to study biological systems. Different models have been combined in a unified framework to capture the complex behavior of biological phenomena. Recently, the hybrid systems paradigm has been applied to various biochemical networks with switching behavior. On the other hand, the main difficult in biological model validation is the precision of predictions provided by the model, in comparison with experimental data. Quantitative information on kinetic parameters is thus hard to obtain and, in addition, experimental data are often of qualitative nature, which poses a major problem to traditional numerical methods. In this work, we try to address this problem by extending the numerical approach to systems with uncertain parameters.

More concretely, we exploit the ability of handling uncertainty of the reachability computation techniques for hybrid systems, a mathematical model for describing the interaction between a discrete and a continuous process. Roughly speaking, the goal of reachability analysis is to study the set of all possible trajectories of a dynamical system. Well-established results for piecewise linear systems are available and they have been successfully applied to numerous biological systems (for example \cite{Batt2005,Jong2008,DangLeguernicMaler2011,GrosuBFGGSB2011}). Nevertheless, nonlinear systems still remain a challenge.

In particular we focus the following computation problem: given a set of initial states in $\real^n$, compute the trajectory set of a parametric discrete-time dynamical system described by the following difference equation:
$\vx(k+1) = \polynom(\vx(k), \vp)$ where $\polynom: \real^n \times \real^m \to \real^n$ is a multivariate polynomial. This problem can be seen as an extension of numerical integration, that is, solving the equations of the dynamics with sets, that is  $x(k)$ and $x(k+1)$ in the equation are subsets of $\real^n$ (while they are points if we only need a single solution, as in traditional numerical integration).

Our interest in polynomial systems is motivated by their applicability in modeling a variety of phenomena in bio-chemical networks. To handle continuous-time models, a time discretization can be needed and conservativeness of approximation needs to be guaranteed (which is the topic of our current research). It is however important to note that discrete-time systems can also be directly used to model many biological systems, since experimental data are often obtained by sampling the  biochemical reaction outputs, and in addition, such models can be readily used for computer aided analysis and numerical simulation.

This problem for polynomial systems was previously considered in the work~\cite{Dang2006,DangSalinas2009}, which was inspired by the techniques from Computer Aided Geometric Design (CADG). In this paper, we pursue the direction by exploiting further the special properties of a technique from CADG, namely the B{\'e}zier simplex and Bernstein expansion, we only need to solve linear programming (LP) problems instead of polynomial optimization problems. In this paper, with a view to biological models, we propose an extension of these techniques to systems with uncertain parameters. We also describe a new techniques specialized for multi-affine systems 

The paper is organized as follows. In Section~\ref{sec:preli} we introduce basic definitions of set integration. We then focus on the problem of computing the image of a set by a parametric polynomial and propose two methods: one is an extension to parametric systems of the method using the Bernstein expansion, and the other method is based on a property of multi-affine functions. Some experimental results on two biological systems are reported in Section~\ref{sec:experiment}: one is a model of cardiac abnormalities, the other is a model of insect nest-site choice.

\section{Set integration}\label{sec:preli}
Let $\real$ denote the set of reals. Throughout the paper, vectors are often written using bold letters. Exceptionally, scalar elements of multi-indices, introduced later, are written using bold letters. Given a vector $\vx$, $x_i$ denotes its $i^{th}$ component. 

We use $\uBox$ to denote the unit box anchored at the origin, that is $\uBox=[0,1]^n$. We use $\polynom$ to denote a vector of $n$ functions such that for all $i \in \{1, \ldots, n \}$, $\polynom_i$ is an $n$-variate polynomial of the form $\polynom_i: \real^n \to \real$. 

We consider a parametric discrete-time dynamical system:
\begin{eqnarray}\label{eq:system}
\vx(k+1) = \polynom(\vx(k), \vp)
\end{eqnarray}
where $\vx \in \real^n$ is the state variables, $\vp \in P \subseteq \real^m$ is the vector of uncertain parameters. The set $P$ is called the parameter set. We assume that $P$ is a convex polyhedron. The initial state $\vx(0)$ is inside some set $X_0 \subset \real^n$, and $X_0$ is called the initial set.

Given a set $X \subset \real^n$, the image of $X$ by $\polynom$ with all possible values of the parameters, denoted by $\polynom(X)$,  is defined as follows: $\polynom(X) = \{ (\polynom_1(\vx, \vp), \ldots, \polynom_n(\vx, \vp))~|~ \vx \in X, \vp \in P \}$. 

When starting from a set of initial states, the dynamical system~(\ref{eq:system}) generates a set of solutions, and at each step the set of all visited states is the image by $\polynom$ of the previous set. We thus focus on the problem of computing the image of a polyhedron $X$ by the polynomial $\polynom$. 

One way to charaterize this set of solutions is to use special convex polyhedra with fixed geometric form, called template polyhedra~\cite{SriramVMCAI05,ChenMWC09}. Indeed, by choosing the templates one can control both the precision and the geometric complexity of the approximations. A template is a set of linear functions defined by an $l \times n$ matrix $H$ and a real-valued vector $\vc \in \real^l$. A template polyhedron is then defined by considering the conjunction of the linear inequalities of the form $\langle H, \vc \rangle = \{ \vx ~|~\bigwedge_{i=1, \ldots, l} H^i \vx \le c_i \}$. The vector $\vc$ is called a polyhedral coefficient vector. 

The template matrix $H$ is assumed to be given; to over-approximate the trajectory set, the polyhedral coefficient vector $\vc \in \real^l$ must satisfy
\begin{equation}\label{eq:surappr-image}
\polynom(X) \subseteq \langle H, \vc \rangle.
\end{equation}

To determine the coefficients $\vc = (c_1, \ldots, c_n)^T$ so that the template polyhedron $\langle H, \vc \rangle$ over-approximate $\polynom(X)$, we can formulate the following optimization problems:
\begin{eqnarray}\label{eq:coef0}
~&\forall i \in \{1, \ldots, l\}, \displaystyle{c_i = \max (\Sigma_{k=1}^n H^i_k \polynom_k(\vx))}\\
~&\text{subj.\ to}~ \vx \in X.
\end{eqnarray}
where $H^i$ is the $i^{th}$ row of the matrix $H$ and $H^i_k$ is its $k^{th}$ element. Note that the above functions to optimize are polynomials. This polynomial optimization problem is computationally difficult (see for example~\cite{Parrilo2006} and references therein). We propose two alternative solutions:
\begin{itemize}
\item Use a linear relaxation of the above optimization problems, so that we can take advantage of well-developed linear programming techniques~\cite{Boyd2004}. Indeed, the Bernstein expansion can be used to compute affine bound functions of polynomials, as shown in the next section.
\item If $\polynom$ is multi-affine, we can exploit its particular properties that allows using corner analysis to reconstruct the set of all trajectories.
\end{itemize}

In the following section, we describe two methods for approximating the image of a box domain by a polynomial, and we then show how to extend these methods to polyhedral domains.

\section{Approximating the image of a box domain}\label{sec:boundBox}

\subsection{Using the Bernstein expansion}
The essence of this method, proposed in~\cite{GarloffLeastSq2008}, for computing an affine bound function is to find a hyperplane that is close to {\em all} the control points, using linear least squares approximation. In this work, we extend this method to parametric polynomial functions. Before presenting the Bernstein expansion, we remark that computing convex lower bound functions for polynomials is a problem of great interest, especially in global optimization. The reader is also referred to~\cite{Garloff2005} for more detail on the approach based on the Bernstein expansion. 

To discuss the Bernstein expansion of polynomials, we use multi-indices of the form $\vvi = (\vvi_1, \vvi_2, \ldots, \vvi_n)$ where each $\vvi_j$ is a non-negative integer. Given two multi-indices $\vvi$ and $\vvd$, we write $\vvi \le \vvd$ if for all $j \in \{1, \ldots, n \}$, $\vvi_j \le \vvd_j$. Also, we write $\frac{\vvi}{\vvd}$ for $\displaystyle (\vvi_1/\vvd_1, \vvi_2/\vvd_2, \ldots, \vvi_n/\vvd_n)$ and $\displaystyle \binom{\vvi}{\vvd}$ for the product of binomial coefficients $\displaystyle \binom{\vvi_1}{\vvd_1} \binom{\vvi_2}{\vvd_2} \ldots \binom{\vvi_n}{\vvd_n}$.

The parametric polynomial $\polynom: \real^n \times \real^m \to \real^n$ can be represented using the power base as follows:
$$\polynom(\vx, \vp) = \sum_{\vvi \in I_{\vvd}} \vva_{\vvi} (\vp)  \vx^{\vvi}$$
where $\vva_{\vvi}$ is a function $\real^m \to \real$; $\vvi$ and $\vvd$ are two multi-indices of size $n$ such that $\vvi \le \vvd$; $I_{\vvd}$ is the set of \emph{all} multi-indices $\vvi \le \vvd$, that is $I_{\vvd} = \{ \vvi ~|~ \vvi \le \vvd \}$; $\vx^{\vvi}$ represents the monomial $ x_1^{\vvi_1}x_2^{\vvi_2}\cdots x_n^{\vvi_n}$. The multi-index $\vvd$ is called the {\em degree} of $\polynom$.

The polynomial $\polynom$ can also be represented using the Bernstein expansion. For  $\vx = (x_1, \ldots, x_n) \in \real^n$, the $\vvi^{th}$ Bernstein polynomial of degree $\vvd$ is defined as follows:
$$\Bern_{\vvd,\vvi}(\vx) = \bern_{\vvd_1,\vvi_1}(x_1) \ldots \bern_{\vvd_n,\vvi_n}(x_n)$$
where for a real number $y$, $\bern_{\vvd_j,\vvi_j}(y) = \binom{\vvd_j}{\vvi_j} y^{\vvi_j} (1 - y^{\vvd_j - \vvi_j})$.

Then, for all $\vx \in \uBox=[0,1]^n$, the polynomial $\polynom$ can be written using the Bernstein expansion as follows: $$\polynom(\vx) = \sum_{\vvi \in I_{\vvd}} \vvb_{\vvi} (\vp) \Bern_{\vvd,\vvi}(\vx)$$
where for each $\vvi \in I_{\vvd}$ the Bernstein coefficient $\vvb_{\vvi}(\vp)$ is defined as:
\begin{equation}
\label{eq:bern}
\vvb_{\vvi}(\vp) = \sum_{\vvj \le \vvi}\frac{\binom{\vvi}{\vvj}}{\binom{\vvd}{\vvj}} \vva_{\vvj} (\vp).
\end{equation}
With respect to our problem of computing the image of a set by a polynomial, the Bernstein expansion is of particular interest since their coefficients allow to derive useful geometric properties. We now show that these coefficients can be used to efficiently compute an affine approximation of the polynomial $\polynom$. 

We denote by $I_{\vvd} = \{ \vvi^j ~|~ 1 \le j \le n_b \}$ be the set of all the multi-indices, $n_b$ is thus their number. The set of all control points are denoted similarly. Let $A$ be a matrix of size $n_b \times (n+1)$ ($n$ is the number of state variables of the dynamical systems in question) such that its elements are defined as follows. For all $1 \le j \le n_b$ and $1 \le k \le n$,
$$
A^j_k = \frac{\vvi^j_k}{\vvd_k}
$$
and $A^j_{n+1} = 1$. 

We pick the centroid point $\vp_c$ in the parameter set $P$. Let $\zeta$ be the solution of the following linear least squares approximation problem:
$$A^T A \zeta = A^T \vvb(\vp_c).$$
Then, the affine function
$$\displaystyle{\tilde{l}(\vx) = \sum_{k=1}^n \zeta_k \vx_k + \zeta_{n+1}}$$
corresponds to the "median" axis of the convex hull of all the control points. It thus suffices to shift it downward by the amount:
\begin{equation}
\label{eq:delta}
\delta = max \{ \tilde{l}(\frac{\vvi^j}{\vvd}) - \vvb^j(\vp) ~|~ 0 \le j \le n_b ~\land~ \vp \in P \}.
\end{equation}

If the function $\polynom$ is linear in the parameters $\vp$, then the above optimization problem is a set of linear programs and can thus be solved efficiently.
This results in a lower bound function
$$l(\vx) = \tilde{l}(\vx) - \delta, ~\text{for all}~ \vx \in \uBox.$$


We now show how the above affine bound functions can be used to solve the optimization problems~(\ref{eq:coef0}) in order to determine the coefficients of a template polyhedron over-approximating the reachable set. The functions to optimize in~(\ref{eq:coef0}) can be seen as the compositions of polynomials $\pi_k$. Since every coefficient $H^i_k$ is constant, each
$$\s^i(\vx) = \Sigma_{k=1}^n H^i_k \pi_k(\vx)$$
is simply a polynomial and we can compute its bound functions. The template polyhedral coefficients can hence be computed by solving the following optimization problems:
\begin{eqnarray}\label{eq:coef2}
\forall i \in \{1, \ldots, l\}, \, c_i = max (\s^i(\vx)) ~\text{subj.\ to}~ \vx \in X.
\end{eqnarray}

\paragraph{Example.}
\begin{figure}[!h]
  \begin{center}
    \scalebox {0.7} {
      \includegraphics{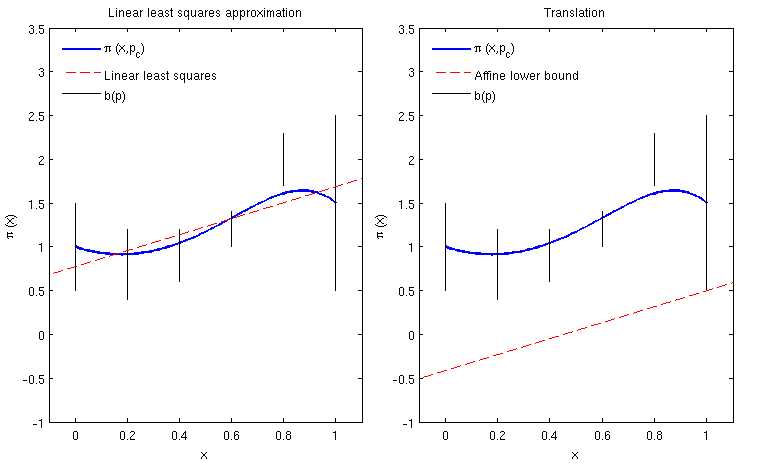}
    }
    \caption{\label{fig:bernstein_llsbounds} Computation of the affine lower bound function using linear least square approximation; the vertical line segments represent the intervals of the Bernstein coefficient values.} 
  \end{center}
\end{figure}
We illustrate the method for computing the bound functions with an example. We consider a  one dimensional polynomial  of degree $5$:
\begin{equation}
 \polynom(\vx) = \vp(1-\vx + 2\vx^4)  + 3\vx^2 - \vx^3 - 2.5\vx^5 \nonumber
\end{equation}
where $\vp$ is a scalar parameter included in $[0.5, 1.5]$.
Using the equation (\ref{eq:bern}), Bernstein coefficients are the following:
$$\vvb(\vp) = ( \vp, 0.8\vp, 0.6\vp +0.3, 0.4\vp+0.8, 0.6\vp+1.4, 2\vp-0.5).$$  
Next we compute the matrix $A$ and the vector $\vvb(\vp_c)$ where the centroid $\vp_c$ is equal to $1$:
$$
A = 
  \left( 
    \begin{array}{cccccc} 
      0 & 0.2 & 0.4 & 0.6 & 0.8 & 1 \\  
      1 & 1 & 1 & 1 & 1 & 1  \\
    \end{array} 
  \right),
$$
$$\vvb(\vp_c)  = \vvb(1)= (1, 0.8, 0.9, 1.2, 2, 1.5),$$  
which gives us the following linear system:
$$ 
\left( 
  \begin{array}{cc} 
    2.2 & 3 \\  
    3 & 6  \\
  \end{array} 
\right)
\left( 
  \begin{array}{c} 
    \zeta_1 \\  
    \zeta_2 \\
  \end{array} 
\right)
=
\left( 
  \begin{array}{c} 
    4.34 \\  
    7.40 \\
  \end{array} 
\right).
$$
By solving this linear equality system, we obtain following least square approximation
$$ \tilde{l}(\vx) =\zeta_1 \vx + \zeta_2 = 0.9143 \vx + 0.7762. $$
Then it follows from (\ref{eq:delta}) that $\delta = 1.1905$ and we get the affine lower bound
$$ l(\vx) = \tilde{l}(\vx) - \delta = 0.9143 \vx - 0.4143.$$

The left plot in Figure~\ref{fig:bernstein_llsbounds} shows the linear least square approximation computed for the  polynomial in question. The other one represents the final affine function computed after shifting downward the previous affine function.

\subsection{Using the property of multi-affine functions}
In this section, we propose a method specialized for multi-affine systems, a particular case of polynomial systems. 
Multi-affine systems are systems composed by polynomials which are affine by each of theirs variables, i.e. if $\vvd$ is the degree of an affine system $\polynom$, $\vvd_k \leq 1$ for all $k\in\{1, \ldots, n \}$.
We will exploit the following property of such systems to achieve better computational efficiency.

\begin{theorem}\label{lem:multiaff}
Given a hyper-rectangle $X \se \real^n$, let $V_X$ be the set of its vertices. If $\polynom$ is a multi-affine function, then
$$\polynom(X) \se \conv \{ \polynom(\vvv) ~|~ \vvv \in V_X \}$$
\end{theorem}
To prove the theorem, we first prove an intermediate result.

We denote the line segment connecting two points $\vx$ and $\vx'$ by $s(\vx, \vx')$. 

\begin{lemma}\label{lem:proj}
If two points $\vx$ and $\vx'$ that differ only in one coordinate (that is there exists only one axis $i \in \{1, \ldots, n \}$ such that $x_i \neq x'_i$ and for all other axes $j \neq i$ $x_j = x'_j$), then for any point $\vy$ in the line segment $s(\vx, \vx')$, its image $\polynom(\vy)$ can be written as a linear combination of $\polynom(\vx')$ and $\polynom(\vx)$.
\end{lemma}
{\em Proof.} It is easy to see that any two points in $s(\vx,\vx')$ also differ only in the axis $i$, (that is $s(\vx, \vx')$ is parallel to the axis $x_i$). The $i^{th}$ coordinate of any point $\vy \in s(\vx, \vx')$ can be expressed as $\vy_i = \lambda_1 \vx_i + \lambda_2 \vx'_i$ where $\lambda_1, \lambda_2 \ge 0$ and $\lambda_1 + \lambda_2=1$. In addition, if the domain of the function $\polynom$ is restricted to the line segment $s(\vx,\vx')$, only the variable $x_i$ varies while all the variables remain unchanged and the function $\polynom$ hence becomes linear in $x_i$. It then follows that $\polynom(\vy)$ can be written as a linear combination of $\polynom(\vx')$ and $\polynom(\vx)$. \qed

We proceed with Theorem~\ref{lem:multiaff}. Let $\vx$ is a point inside $X$. Let $\vy$ and $\vz$ be the projections of $\vx$ on two opposite faces of $X$ perpendicular to the axis $x_n$. It follows from Lemma~\ref{lem:proj} that $\polynom(\vx)$ can be written as a linear combination of $\polynom(\vy)$ and $\polynom(\vz)$. Furthermore, $\vy$ can now be treated as a point inside a hyper-rectangle in $(n-1)$ dimensions $(x_1, \ldots, x_{n_1})$, and by successive projections until dimension $x_1$ we can prove that $\polynom(\vx)$ can be written as a linear combination of the vertices of $X$.
\qed

Another proof of this well-known property of multi-affine functions can be found in \cite{Belta2006}

Note that the above theorem is only applicable to the sets which are axis-aligned hyper-rectangles. Hence, even if the initial set satisfies this condition, after the application of $\polynom$, the resulting set is a general polytope and we need to approximate it by an axis-aligned hyper-rectangle. Using the theorem, to compute $\polynom(X)$ it suffices to compute the images of the vertices of $X$ and then take the convex hull of the resulting points.

Before continuing, we remark that a number of different methods for the multi-affine systems have also been developed in \cite{Batt07, Berman07}. These methods are however based on a rectangular partition of the state space, while we allow reachable sets to be represented by unions of polytopes.

\section{Approximating the image of a polyhedral domain}
As mentioned earlier, the above described methods can be applied only when the set $X$ is inside the unit box $\uBox$ anchored at the origin. To extend it to polyhedral domains, we transform the polyhedra to the unit box by two methods: (1) via an (oriented) box approximation, and (2) by rewriting the polynomials using a change of variables.

If we over-approximate $X$ with a box $B$, it is then possible to derive a formula expressing the Bernstein coefficients of $\polynom$ over $B$. However, this formula is complex and its representation and evaluation can become expensive.

We alternatively consider the composition of the polynomial $\polynom$ with an affine transformation $\transf$ that maps the unit box to $B$. The functions resulting from this composition are still polynomials, for which we can compute their bound functions over the unit box, using the formula~(\ref{eq:bern}) of the Bernstein expansion. This is explained more formally in the following.

Let $B$ be the bounding box of the polyhedron $X$, that is, the smallest box that includes $X$. The affine function $\transf$ that maps the unit box $\uBox$ to $B$ can be easily defined as:  $\transf(\vx) = diag(\lambda) \vx + \vg$ where $\vg \in \real^n$ such that $g_i = l_i$, and $diag(\lambda)$ is a $n \times n$ diagonal matrix with the elements on the diagonal defined as follows: for each $i \in \{1, \ldots, n \}$, $\lambda_i = h_i - l_i$. The composition $\comp =(\polynom ~o~ \transf)$ is defined as $\comp(\vx) = \polynom(\transf(\vx))$. Note that $\comp = \polynom ~o~ \transf$. Then, $\polynom(X) \subseteq \comp(\uBox)$.

Since multi-affine functions are not closed under the above composition. They can become quadratic. In this case, we use the following blossoming principle to transform a polynomial to an affine function.

\begin{theorem}[Blossoming principle] For any polynomial $\polynom: \real^n \to \real^m$ of degree $d$, there is a unique symmetric multi-affine map $\blossom: \real^n \times \ldots \times \real^n \to \real^m$ such that for all $\vx \in \real^n$
$$\blossom(\vx, \ldots, \vx)=\polynom(\vx).$$
\end{theorem}
The proof of the above theorem is standard and can be found in~\cite{PrautzschBoehm2002}. We can thus compute the image $\polynom(B)$ by the convex hull of all the images of the vertices of $X$ by the multi-affine function $\blossom$. Nevertheless, it is important to note that when $\vx$ inside an rectangle, the arguments of the blossom varies only on the diagonal of the corresponding rectangle in the augmented space $\real^{nd}$. Thus the convex hull is an over-approximation.

\section{Experimental results}\label{sec:experiment}
We implemented the above described algorithms, and their time efficiency was also evaluated by considering a number of randomly generated polynomials, which shows that the algorithms can handle general polynomial systems in up to $11$ dimensions with reasonable computation time. In the following, we demonstrate their application to two biological systems.

\subsection{A model of insect nest-site choice}
We study a model of a decision making mechanism used by a swarm of honeybees to select one  among many different nest-sites~\cite{Britton2002}. This is built upon classical mathematical models of the spread of diseases, information and beliefs. The bee population is divided into $5$ groups: 
\begin{itemize}
\item $X$: neutral bees that have not chosen a site
\item $Y_1$: evangelic bees dancing for the site $1$
\item $Y_2$: evangelic bees dancing for the site $2$
\item $Z_1$: non-evangelical bees which have been converted to the site $1$
\item $Z_2$: non-evangelical bees which have been converted to the site $2$
\end{itemize}   
The bees dance not only to advertise the quality of a site but also to transmit to the other bees information about the direction and distance to the site. Non-evangelical bees stay idle but may take up dancing when they are stimulated by a dancing bee. For simplicity, we also use the names of the groups to denote the respective numbers of individuals in each group. 
 
The equations describing the evolutions of the variables are as follows:
\begin{eqnarray}
\dot{X}(t) & = & -\beta_1 X(t) Y_1(t) - \beta_2 X(t) Y_2(t), \nonumber\\
\dot{Y}_1(t) & = & \beta_1 X(t) Y_1(t) - \gamma Y_1(t) + \delta \beta_1 Y_1(t) Z_1(t) + \alpha \beta_1 Y_1(t) Z_2(t), \nonumber \\
\dot{Y}_2(t) & = & \beta_2 X(t) Y_2(t) - \gamma Y_2(t) + \delta \beta_2 Y_2(t) Z_2(t) + \alpha \beta_2 Y_2(t) Z_1(t), \nonumber \\
\dot{Z}_1(t) & = & \gamma Y_1(t) - \delta \beta_1 Y_1(t) Z_1(t) - \alpha \beta_2 Y_2(t) Z_1(t), \nonumber \\
\dot{Z}_2(t) & = & \gamma Y_2(t) - \delta \beta_1 Y_2(t)Z_2(t) - \alpha \beta_2 Y_1(t) Z_2(t), \nonumber
\end{eqnarray}
where $\beta_1$ and $\beta_2$ are scalar parameters representing a measure of how vigorously the bees dance for the sites $1$ and $2$ respectively; $\alpha$ is the parameter representing the proportionality of switching back to the neutral state, and $\gamma$ is the proportionality of conversion from the dancing state to the idle state. The proportionality of conversion from the neutral state to any site $Y_i$ is $1$ and from the idle state to $Y_i$ is $\delta$.

In this case study, we want to study the influence of the parameter $\beta_2$ on the possibility of achieving a consensus on nest-site choice.

Using the Euler discretization method we obtains the following discrete-time dynamics. 
\begin{eqnarray}
X(k+1) & = & X(k) + h(-\beta_1 X(k) Y_1(k) - \beta_2 X(k) Y_2(k)), \nonumber\\
Y_1 (k+1)& = & Y_1(k) + h(\beta_1 X(k) Y_1(k) - \gamma Y_1(k) + \delta \beta_1 Y_1(k) Z_1(k) + \alpha \beta_1 Y_1(k) Z_2(k)), \nonumber \\
Y_2(k+1) & = & Y_2(k) + h(\beta_2 X(k) Y_2(k) - \gamma Y_2(k) + \delta \beta_2 Y_2(k) Z_2(k) + \alpha \beta_2 Y_2(k) Z_1(k)), \nonumber \\
Z_1(k+1) & = & Z_1(k) + h(\gamma Y_1(k) - \delta \beta_1 Y_1 Z_1 - \alpha \beta_2 Y_2 Z_1), \nonumber \\
Z_2(k+1) & = & Z_2(k) + h(\gamma Y_2(k) - \delta \beta_1 Y_2(k) Z_2(k) - \alpha \beta_2 Y_1(k) Z_2(k)), \nonumber
\end{eqnarray}
We choose a discretization time $h=0.01$. In the following, $\beta_1 N = 1.0$, the initial set is a small box centered at $(N,0,0,0,0)$, where $N$ is the total number of honeybees arbitrary fixed to $1000$. 

\begin{figure}[!h]
\centering
 \includegraphics[scale=.5]{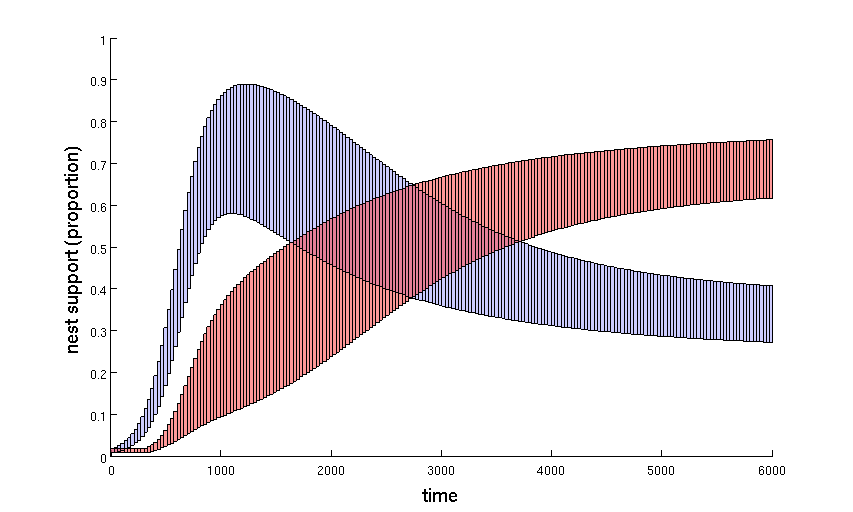}
\caption{\label{fig:noconsens} No consensus is observed with $\alpha = 0.7, \gamma = 0.3$ and $\beta_2 \in [1,1.2]$.}
\end{figure}

\begin{figure}[!h]
\centering
 \includegraphics[scale=.5]{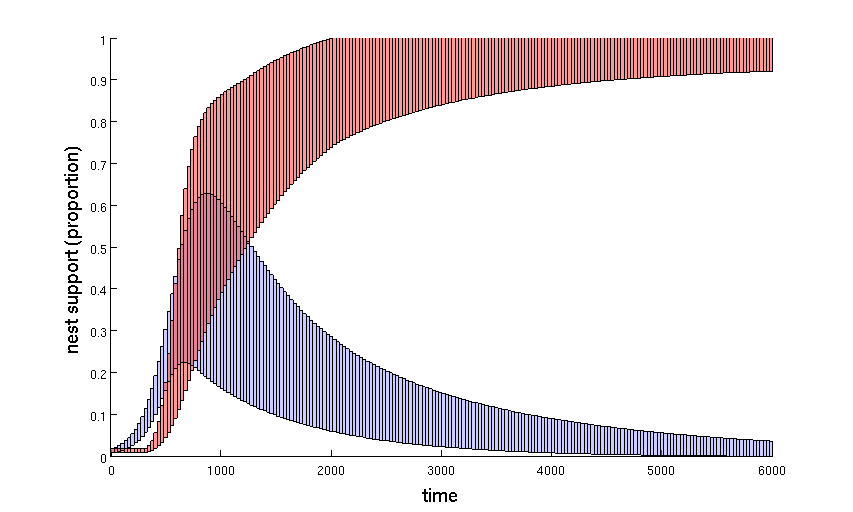}
\caption{\label{fig:consens}A consensus for the site $2$ is observed with $\alpha = 0.7, \gamma = 0.3$ and $ \beta_2 \in [1.5,2]$.}
\end{figure}

Figure~\ref{fig:noconsens} and Figure~\ref{fig:consens} show the evolution of the proportions of converted honeybees ($(Y_i + Z_i)/N$) during $6000$ iterations, with different interval values of the parameter $\beta_2$. The blue sets (in lighter color) correspond to the proportions of bees supporting the site $1$ and the red sets (in darker color) correspond to those supporting the site $2$. These results were obtained using the method for multi-affine systems which took about $7.5s$

First we consider that at the beginning no honeybees dance for the site $2$, in other words $\beta_2=0$. After $300$ iterations the site $2$ is discovered and its propaganda begins. We can observe two main comportments: a consensus and the absence of consensus. We consider that there is a consensus if the distance between the proportions of honeybees supporting the site $1$ or $2$ is large and tends to increase.

Figure~\ref{fig:noconsens} shows the analysis using the parameters $\alpha = 0.7, \gamma = 0.3$ and $ \beta_2 \in [1,1.2]$. We observe that $\beta_2 $ is marginally superior to  $\beta_1$, however we can see that there is no consensus between the two evangelical groups. 

Figure~\ref{fig:consens} shows that a consensus can be observed if the measure of how vigorously the bees dance for sites $2$ is increased. In this analysis the parameters stay the same except for  $\beta_2 \in [1.5,2.0]$. We observe a clear consensus for the site $2$ despite its lateness.

\begin{figure}[!h]
\centering
 \includegraphics[scale=.5]{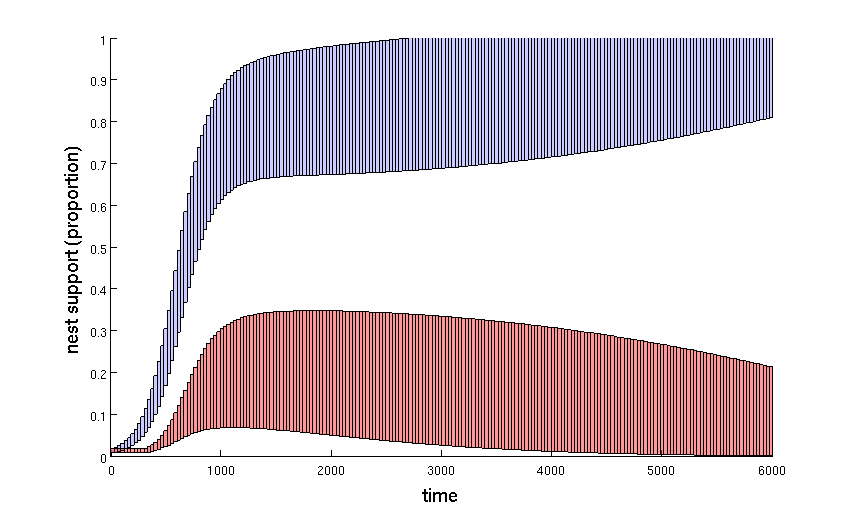}
\caption{\label{fig:consens2}A consensus for the site $1$ is observed with $\alpha = 0.2, \gamma = 0.3$ and $\beta_2 \in [1,1.2]$.}
\end{figure}

Another way to obtain a consensus without modifying $\beta_2$ is to reduce $\alpha$ which corresponds to the proportionality of honeybees reconversion to another site. When $\alpha = 0.2$, the late propaganda of the site $2$ leads to a consensus to choose the site $1$ as shown in Figure~\ref{fig:consens2}.

\begin{figure}[!h]
\centering
\includegraphics[scale=.35]{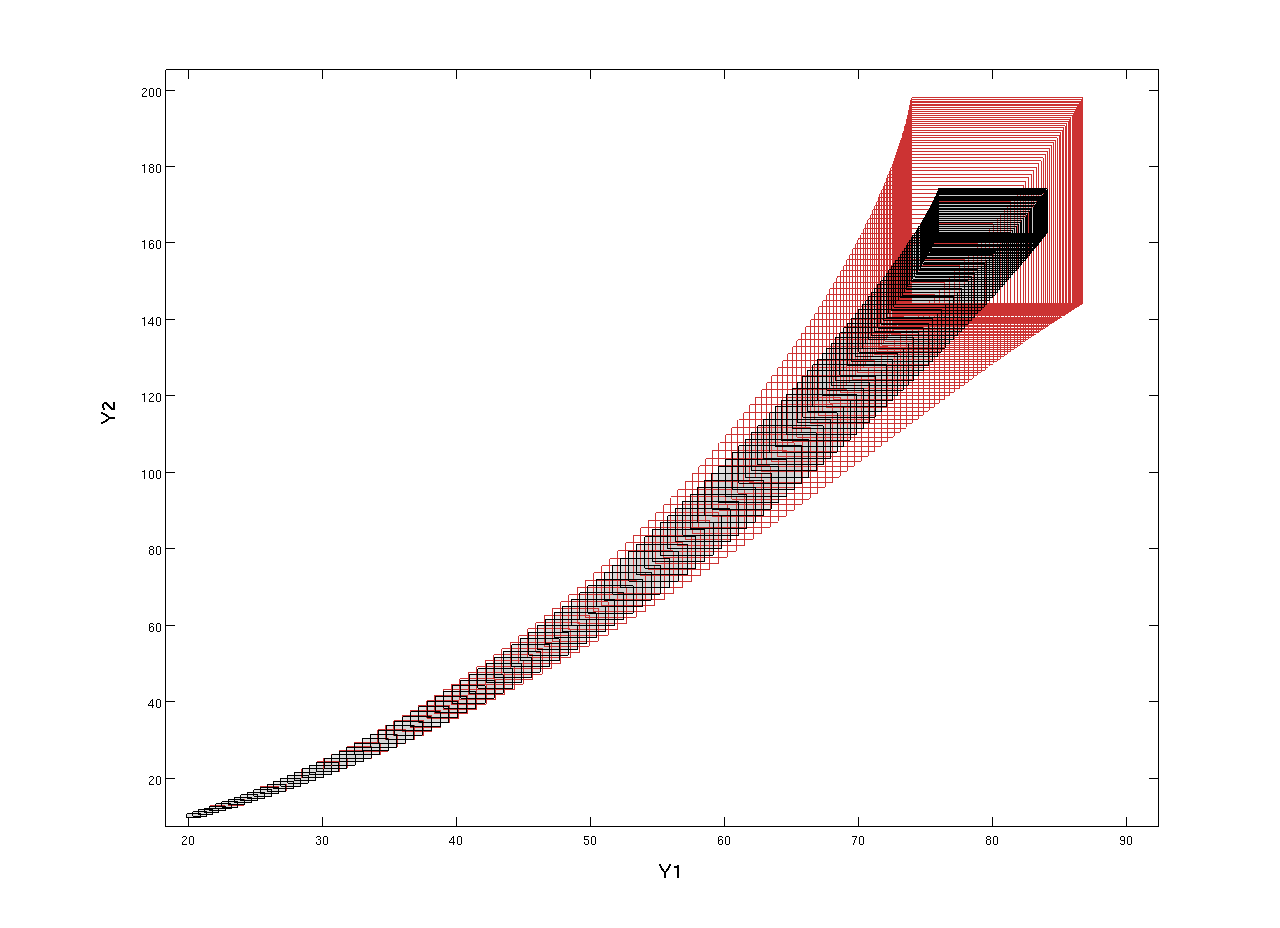}
\caption{\label{fig:methods_precision} Reachability analysis during $150$ iterations using the two reachability methods, the black outlines set are computed with the Bernstein expansion method, the red outlines set correspond to the result using the method for multi-affine systems.}
\end{figure}

Finally, to compare the two proposed methods, Figure~\ref{fig:methods_precision} shows the projection of the reachable sets on the variables $Y_1$ and $Y_2$ during the first $150$ iterations, with $\alpha = 0.7$, $\gamma = 0.3$ and $ \beta_2 = 1.2$. We plot the results obtained by the two methods and we can see that the sets computed using the method for multi-affine systems are larger than those computed using the Bernstein expansion. We observe a gain of precision with the Bernstein method, however the computation time of this method is $25.06s$ and is more than $250$ times superior than the computation time of the other method which is $0.09s$. 

\subsection{A model of cardiac abnormalities}

The second case study comes from a simplified model of cardiac cells~\cite{GrosuBFGGSB2011} which describes the conditions under which cardiac abnormalities (such as, ventricular tachycardia and fibrillation) can arise. The behavior of cardiac cells can be modelled using a Detailed Ionic Model (DIM) which are differential equation models of reaction-diffusion type. Such models are complex with a lot of parameters. An alternative model is called Minimal Resistor Model (MRM), the parameters of which can be identified from either experimental data or from DIM-based simulation results. The main advantage of the MRM model is that it exhibits more accurate mesoescopic behavior and additionally allows faster simulations. From the MRM model, the authors of~\cite{GrosuBFGGSB2011} derived an equivalent Minimal Conductance Model (MCM) which is exactly in form of a genetic regulatory network model (GRM). By approximating slow sigmoidal switches in the MCM model with a succession of ramps, a Hybrid Automaton (\cite{Hen96}) is obtained. 

In this case study, we consider the Hybrid Automaton model and the following problem: find the parameter ranges which correspond to the behaviors where the cardiac cells lose their excitability. These ranges can then be used to infer the corresponding parameter ranges in the DIM model. The hybrid automaton has two modes, denoted by $loc1$ and $loc2$.
\begin{itemize}
   \item  $loc1$:
     \begin{eqnarray*}
     \frac{du}{dt} &= e - g_1 u \\
     \frac{dv}{dt} &= (1-v) g_1
     \end{eqnarray*}
   \item  loc2:
     \begin{eqnarray*}
     \frac{du}{dt} &= e - g_2 u \\
     \frac{dv}{dt} &= -v g_2;
     \end{eqnarray*}
\end{itemize}
The transition guard from $loc1$ to $loc2$ is $u \ge 0.06$. In this model, an \emph{action potential} is a change in cell transmembrane potential $u$ as a response to an external stimulus (current) $e$ defined by $e = 0.66$ if $t < 0.25$ and $e = 0$ if $t  \ge 0.25$. The initial state corresponds to
$u=0$. In the above equations, $g_1$ and $g_2$ are parameters such that $1<g_1<180$ and $1<g_2<10$. The safety property we are interested in is stated as follows: the system, when at location $loc2$, will never reach a state where $u \ge 0.13$. To this end, we add a fictitious location $loc3$ and let the transition guard from $loc2$ to $loc3$ be $u \ge 0.13$. We also consider the two parameters as two additional variables with derivatives equal to $0$. The problem is now to compute a set of possible values of $g_1$ and $g_2$ which guarantee that the system does not reach $loc3$.

\begin{figure}[!h]
\centering
\includegraphics[scale=0.3]{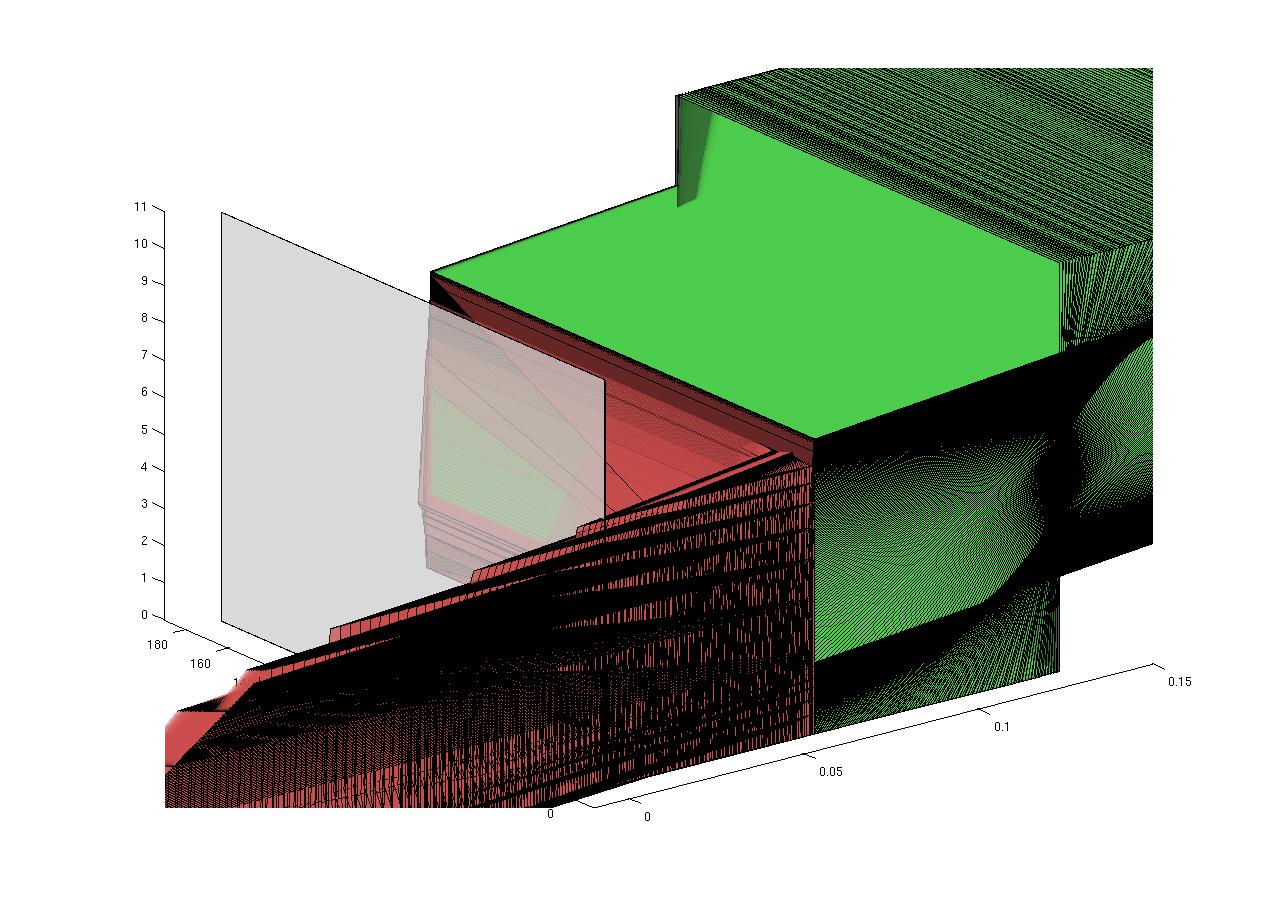}
\caption{\label{fig:3state} Backward reachability analysis, the change of color represents the change of location.}
\end{figure}

The model hence becomes a Hybrid Automaton with $4$ variables and their dynamics are described by multi-affine differential equations, which we discretize in time.  Using backward reachability analysis starting from the unsafe set, we can over-approximate the set of $(g_1,g_2)$ for which the system can reach the guard of the transition to $loc3$. The figure~\ref{fig:3state} shows the projection on the variables $u$, $g_1$, $g_2$ of the reachable set, starting from the guard leading to the location $loc3$. Its intersection with the hyperplane defined by $u=0$ (in grey) corresponds to the set of unsafe values of $g_1$ and $g_2$. This computation uses dynamical templates and static boxes, and the computation time is $1.52s$. The complement of this set is the set of safe parameter values.

\begin{figure}[!htb]
\centering
\includegraphics[scale=0.3]{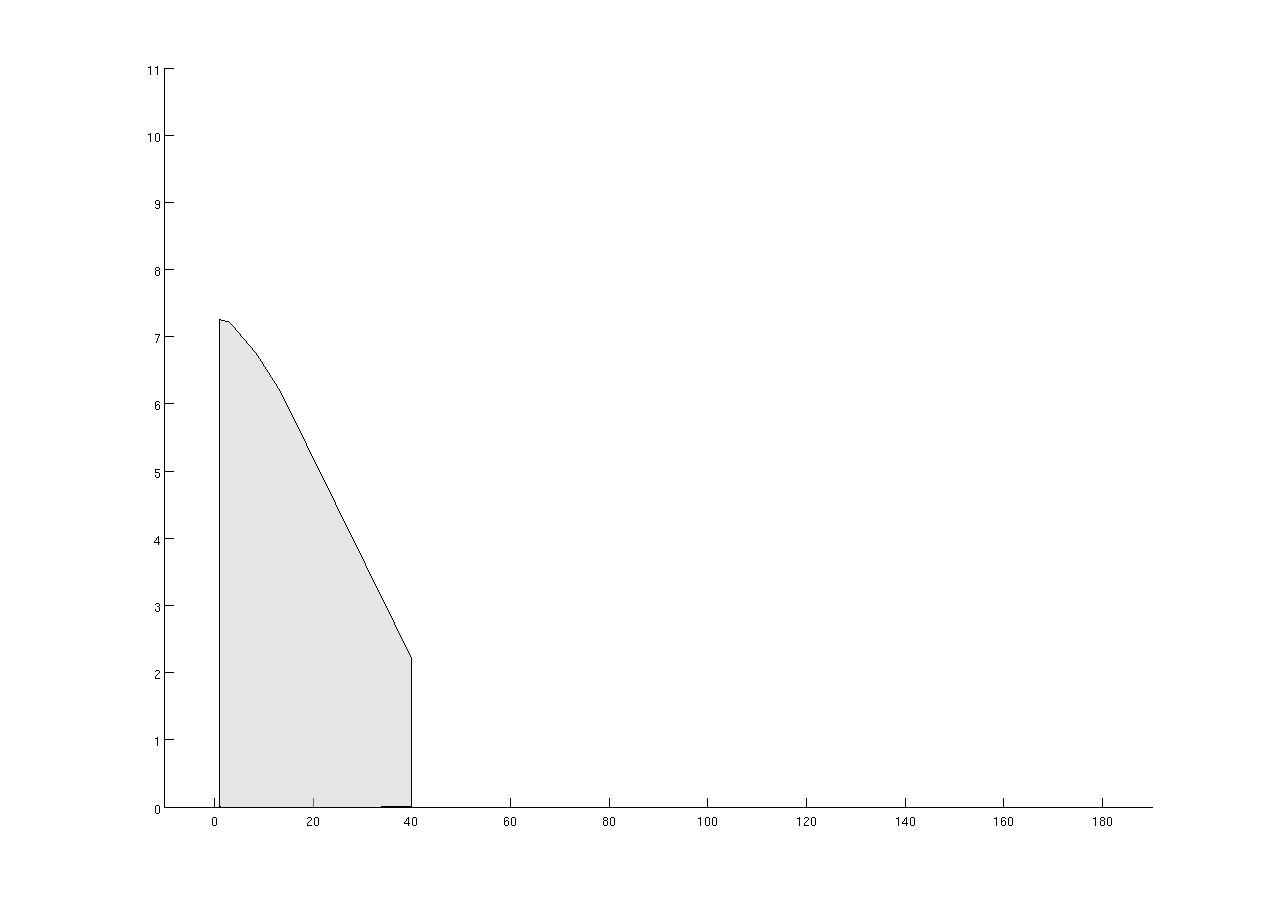}
\caption{\label{fig:3stateRes} Over-approximation of the parameter values $(g_1,g_2)$ that lead to the unsafe location.}
\end{figure}

\section{Related work}
Our reachability analysis approach is similar to a number of existing ones for continuous and hybrid systems in the use of linear approximation. Its novelty resides in the efficient way of computing linear approximations. Indeed, a common method to approximate a non-linear function by a piecewise linear one, as in the hybridization approach~\cite{AsarinDangGirard2007,DT11} for hybrid systems, requires non-linear optimization. 

Besides constrained global optimization, other important applications of the Bernstein expansion include various control problems~\cite{Garloff99} (in particular, robust control). The approximation of the range of a multivariate polynomial over a box and a polyhedron is also used in program analysis and optimization (for example~\cite{Tchoupaeva04,Clauss04}). In the hybrid systems verification, polynomial optimization is used to compute barrier certificates \cite{PrajnaJadbabaie04}. 

\section{Conclusion}

The reachability computation methods we proposed in this paper combine the ideas from optimization and the properties of polynomials. These results can be readily applicable to hybrid systems with polynomial continuous dynamics. We are currently working on the problem of parameter synthesis using the analysis techniques from robust control. The next step will be a more intensive evaluation of the methods on biological systems.

\bibliographystyle{eptcs}
\bibliography{hybcompute,mybiblio}

\end{document}